\documentclass[12pt]{spie}   
\usepackage{amsmath,amsfonts,amssymb}
\usepackage{graphicx} 
\usepackage{setspace}
\usepackage{tocloft}
\usepackage{caption}
\usepackage{subcaption}
\usepackage{fancyref}
\usepackage{tabularx}
\usepackage{multirow} 
\usepackage{amsmath,amsfonts,amssymb}
\usepackage{graphicx}
\usepackage{float}
\usepackage{wrapfig}
\usepackage{gensymb}
\usepackage[colorlinks=true, allcolors=blue]{hyperref}

\newcolumntype{Y}{>{\centering\arraybackslash}X}

\title{Characterization of diamond-turned optics for SCALES}

\author[a,*]{Isabel J. Kain}
\author[a.c]{Phil Hinz}
\author[d]{Marius Doetz}
\author[d]{Benjamin Bulla}
\author[c]{Renate Kupke}
\author[c]{Daren Dillon}    
\author[a]{Andrew Skemer}
\author[c]{Deno Stelter}
\author[c]{Michael Gonzales}
\author[c]{Nicholas MacDonald}
\author[b]{Aditi Gangadharan}
\author[c]{Cristian Rodriguez}
\author[c]{Christopher Ratliff}
\author[e]{Mackenzie R. Lach}
\author[e]{Steph Sallum}

\affil[a]{University of California Santa Cruz (UCSC) Department of Astronomy and Astrophysics}
\affil[b]{University of California Santa Cruz (UCSC) Department of Physics}
\affil[c]{University of California Observatories (UCO)}
\affil[d]{son-x GmbH}
\affil[e]{University of California Irvine (UCI) Department of Physics and Astronomy}

\cftpagenumbersoff{figure}
\cftpagenumbersoff{table} 
\begin{document} 

\authorinfo{*Corresponding author information: \\Isabel J. Kain, ijkain@ucsc.edu}{}{}

\maketitle
\begin{abstract}

High-contrast imaging has been used to discover and characterize dozens of exoplanets to date. The primary limiting performance factor for these instruments is contrast, the ratio of exoplanet to host star brightness that an instrument can successfully resolve. Contrast is largely determined by wavefront error, consisting of uncorrected atmospheric turbulence and optical aberrations downstream of AO correction. Single-point diamond turning allows for high-precision optics to be manufactured for use in astronomical instrumentation, presenting a cheaper and more versatile alternative to conventional glass polishing. This work presents measurements of wavefront error for diamond-turned aluminum optics in the Slicer Combined with an Array of Lenslets for Exoplanet Spectroscopy (SCALES) instrument, a 2-5 micron coronagraphic integral field spectrograph under construction for Keck Observatory. Wavefront error measurements for these optics are used to simulate SCALES’ point spread function using physical optics propagation software \texttt{poppy}, showing that SCALES' contrast performance is not limited by wavefront error from internal instrument optics.

\end{abstract}


\begin{spacing}{1.0}   

\section{Introduction}
\label{sect:intro}

The Slicer Combined with an Array of Lenslets for Exoplanet Spectroscopy (SCALES) instrument is an infrared ($2-5 \mu m$) integral field spectrograph being built for the Keck II Adaptive Optics system. Designed for high-contrast imaging of exoplanets, SCALES will be able to directly spectroscopically characterize the atmospheres of exoplanets and brown dwarfs as cold as $300$K. These cold objects represent an older demographic of exoplanets compared to the few dozen planets characterized using existing high-contrast imaging instruments, which operate at shorter wavelengths and are limited to young giant planets which are self-luminous with residual heat from recent formation.

High-contrast imaging involves suppressing starlight from the planet's host, which outshines the planet itself by many orders of magnitude. The intensity of the stellar halo (and thus the best achievable contrast) is driven by the wavefront error (WFE) of the incoming signal. WFE can be imparted by atmospheric turbulence (which can be compensated by adaptive optics) and by aberrations in the optical surfaces within a telescope and instrument. Optical aberrations may create quasi-static artifacts in the PSF that are difficult to correct for (and may even look identical to planets), degrading performance and decreasing an instrument's sensitivity to high-contrast targets. Characterization of optics is crucial for understanding an instrument's eventual sensitivity.

\begin{table}[H]
\centering
\renewcommand{\arraystretch}{1.15}
\caption{\label{tab:specs} Summary of foreoptics specifications (specified for 293$\degree$K), including flat fold mirrors (FMs), off-axis parabolas (OAPs), and an off-axis ellipse (OAE). All optics are manufactured by son-x.}
\begin{tabularx}{0.95\textwidth}{ c | Y | Y | Y | c | Y }
    \hline\hline
     Optic &  Mechanical aperture diameter & Center thickness & Parent radius of curvature & Conic & Off-axis distance \\
         & [mm] & [mm] & [mm] &    & [mm] \\ 
     \hline\hline
        OAP1.1 & 35.14 & 28.01 & 401.60 $\pm$ 0.1\% & -1.0 & 52.87 $\pm$ 0.2 \\
        OAP1.2 & 50.19 & 34.52 & 803.10 $\pm$ 0.25\% & -1.0 & 220.85 $\pm$ 0.2 \\
        OAE    & 40.15 & 28.52 & 391.27 $\pm$ 0.25\% & -0.7 & 61.76 $\pm$ 0.2 \\
        FM1    & 30.12 & 16.00 & -- & -- & -- \\
        FM2    & 45.17 & 21.00 & -- & -- & -- \\
        FM3    & 15.33 & 10.50 & -- & -- & -- \\
        FM4    & 60.23 & 15.00 & -- & -- & -- \\
        FM5    & 90.35 & 21.00 & -- & -- & -- \\
    \hline\hline
\end{tabularx}
\end{table}

Characterization of SCALES optics is also motivated by our use of diamond turned aluminum mirrors, a technique which has not historically had comparable performance to conventional polishing for use in astronomical instruments. The SCALES instrument is designed using a 6061-T651 bench and mounts, when possible.  Optics are constructed of RSA 6061 aluminum to produce an instrument of near uniform coefficient of thermal expansion (CTE).  By using diamond turned aluminum substrate optics the instrument can be aligned at room temperature and maintain its alignment when cooled to the 100k operating temperature.  Bare aluminum is difficult to polish to an optical quality appropriate for astronomical instrumentation given the softness of the material, though polishing recipes exist which trade excellent surface finish for less dependable surface form \cite{Maloney2016EveryMRF}; instead, single-point diamond turning can be used, where the surface of the optic is turned into the desired shape with a diamond-tipped cutting tool. This technique is useful for making aspheric optics, though tooling grooves left over from manufacturing may impart unwanted scattering and diffraction effects.


\begin{wrapfigure}[31]{r}{0.25\textwidth}
    \centering
    \vspace{-\intextsep}
    \includegraphics[width=0.9\linewidth]{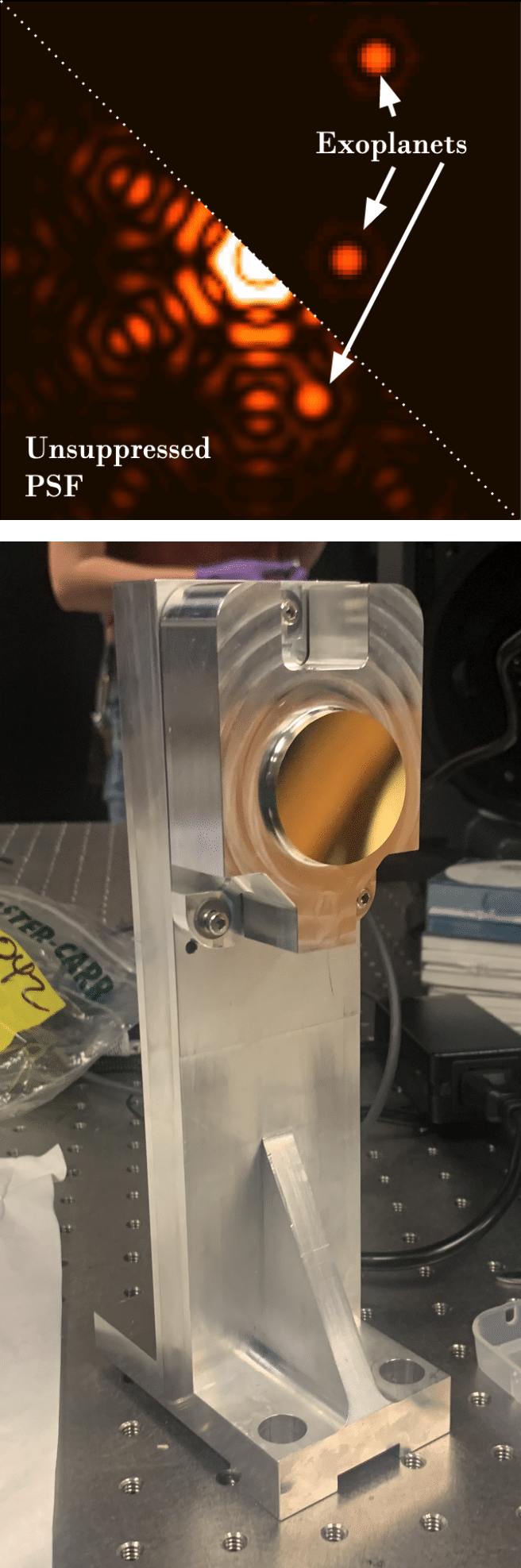}
    \caption{Top: Optical aberrations can create speckles, which can be indistinguishable from exoplanets. Bottom: OAP1.1 during warm testing, fixed to a low-stress mount designed and built at UCO. } 
\end{wrapfigure}

\begin{figure}[ht]
    \centering
    \includegraphics[width=0.9\textwidth]{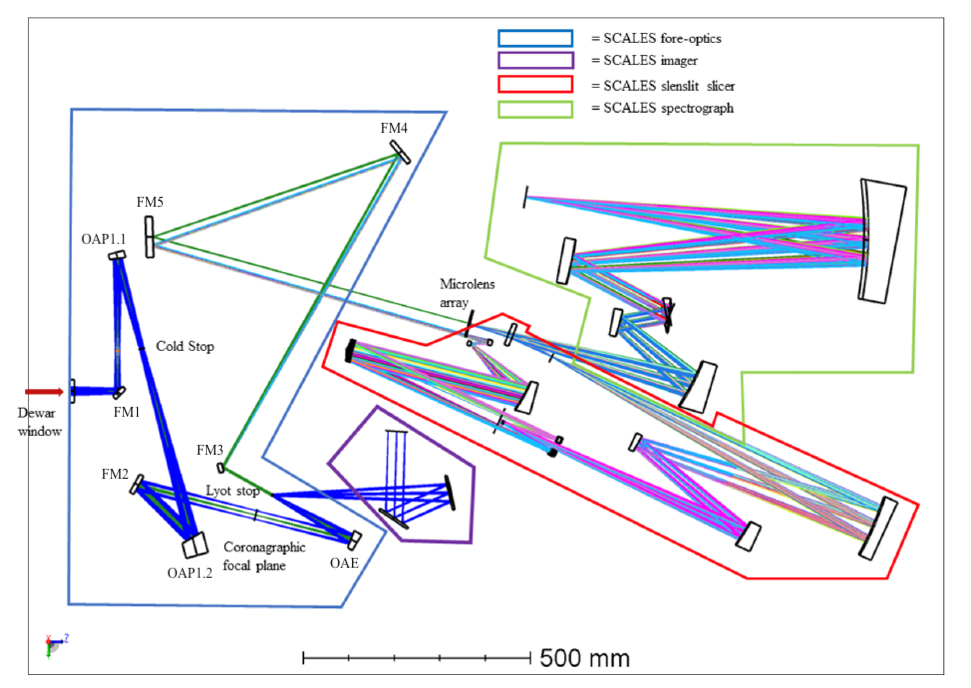}
    \caption{Labeled optical diagram of SCALES from Kupke et al. 2022 \cite{Kupke2022SCALESDesign}. SCALES accepts an f/35 beam from the Keck adaptive optics system, which then passes through the foreoptics and forms an image on the lenslet array. The tip-tilt mechanism steers the beam between the low-resolution IFU (where the beam passes through the lenslet array and into the spectrograph) and the medium-resolution IFU (where the beam passes through a small section of the lenslet array, is sent through a slicer, then is sent through the spectrograph). The eight foreoptics, which are measured in this work, are outlined in blue.  
    }
    \label{fig:optics}
\end{figure}

The overall SCALES instrument design is presented in Skemer et al. 2022 \cite{Skemer2022DesignObservatory}, and the optical design is presented in Kupke et al. 2022 \cite{Kupke2022SCALESDesign}. SCALES comprises four major optical subsystems, which are annotated in Figure \ref{fig:optics}: a set of foreoptics that relay the input beam into the spectrograph, an integral field spectrograph with low-resolution (R$\sim$35-200) and medium-resolution (R$\sim$3,000-6,000) modes, and an imaging channel. The foreoptics are fed by the Keck AO system and place a focal plane image on a lenslet array, and a flat optic (FM3) mounted to a tip-tilt mechanism manufactured by Physik Instruments (PI) is used to steer the image between different resolution modes of the IFU. In the low-resolution mode, the grid of micro-pupils from the lenslet array is sent directly through the spectrograph. The medium resolution IFU uses a ”slenslit” architecture, where the lenslet spots are sent first through an image slicer which reformats a nine by nine lens area of the lenslet array into a pseudo-slit which can be dispersed across the full detector width.  The slicer output injects the reformatted beam back into the main spectrograph.


The SCALES wavefront is spatially sampled at the lenslet array, meaning any optical aberrations downstream of the lenslet array won't impact the PSF. Only aberrations of the foreoptics, the optical subsystem between SCALES' entrance window and the lenslet array, have an effect on contrast performance. The optical specifications of these foreoptics are outlined in Table \ref{tab:specs}.

This paper characterizes the low- and high-spatial frequency wavefront error contributions and thermal stability of all received SCALES foreoptics, and simulates the corresponding impact on contrast performance. Section \ref{sect:characterization} presents measurements for surface figure (\ref{subsect:figure}) and surface roughness (\ref{subsect:roughness}) of all received optics, and discusses the rejection of one optic after failing thermal stability testing (\ref{subsect:failure}). Finally, wavefront error measurements are used in Section \ref{sect:contrast} to calculate contrast performance for the SCALES instrument. All measurements are summarized in Section \ref{sect:summary}.

\section{Characterization of diamond-turned optics}
\label{sect:characterization}


Wavefront error covers a range of spatial frequencies, which are divided into different regimes by their effects on overall optical performance. Low-spatial-frequency errors are deviations that span the entire surface of an optic, i.e. the lower-order Zernike terms, which reshape the core of an optical system's point spread function (PSF). Mid-spatial-frequency errors, sometimes referred to as waviness or ripple, scatter light out of the core of the PSF without changing its shape, displacing it instead into the rest of the focal plane image, pushing light into the Airy rings or introducing hazes or patchy illumination several $\lambda$/D from the core of the PSF. High-spatial-frequency errors correspond to small-scale surface texture, and scatter light at large angles out of the beam path.

While spatial frequency is typically specified in cycles per millimeter, its impact on the PSF is better defined by cycles per aperture. For example, a 0.1 cycles per millimeter aberration across a 1 centimeter diameter optic is 1 cycle per aperture, which might look like tip-tilt error. The same spatial frequency across a 2 meter diameter optic is 200 cycles per aperture, and instead scatters light into the outskirts of the PSF. Because the optics presented in this work are of similar sizes (ranging from 30-90mm in diameter), we stick with a conventional definition in terms of cycles per millimeter. This work refers to low-spatial-frequency error as surface figure, and high-spatial-frequency error as surface roughness. We define low spatial frequencies as 0.0018 - 0.45 mm$^{-1}$, mid spatial frequencies as 0.45-1 mm$^{-1}$, and high spatial frequencies as 1-530 mm$^{-1}$, following the sensitivity ranges of our laboratory instruments as per industry convention.




The tolerances for propagated wavefront error through the SCALES foreoptics are detailed in Kupke et al. 2022 \cite{Kupke2022SCALESDesign}. To keep distortions imparted by SCALES foreoptics well below the error level of the Keck AO system, a maximum wavefront error of 24 nm RMS surface figure and 5 nm RMS surface roughness (with a goal of 3 nm RMS surface roughness) for each optic were established.

At the time of submission, we have received two foreoptics (including OAP1.1 and two versions of FM3, one made of RSA 443 and one made of RSA 6061), while six out of the eight foreoptics (OAP1.2, OAE, FM1, FM2, FM4, FM5) are still being machined and coated. OAP1.1 has been received, tested, and accepted. An initial version of FM3 (the RSA 443 substrate version) failed during testing (see Section \ref{subsect:failure}). The optic was remade by son-x with an RSA 6061 substrate, but for the sake of scheduling was integrated directly with the tip-tilt mechanism before we carried out the tests detailed in this work. Testing of the remade RSA 6061 FM3 will run concurrently with testing of the tip-tilt mechanism. The following describes the tests carried out on the optics received so far, and the same test procedure will be applied to the remaining optics as they are received.


\subsection{Surface figure}
\label{subsect:figure}

Low-spatial-frequency error refers to the overall surface figure of an optic, which deforms the point spread function of an optical system. The surface form of SCALES foreoptics must meet specifications at both ambient and operating temperatures, so surface form must remain stable and within specifications through repeat cryogenic cycles. 

All optics are cryo-annealed before machining to relieve internal material stresses and prevent later deformation. Once parts are machined, coated, and shipped to UCSC, surface figure measurements are taken with a Zygo Verifire\texttrademark   Fizeau interferometer, which has a spatial frequency range of 0.0018-0.45 mm$^{-1}$. A baseline measurement of surface form at ambient temperature is taken for each optic using the setup shown in Figure \ref{fig:zygo-direct}. However, the SCALES instrument will undergo dozens of cryocycles within its lifetime, which requires stable surface figure over repeat cryocycles. To measure elastic deformation, the RSA 443 version of FM3 (and, in future work, every other flat optic) was continuously monitored throughout two full cryocycles inside a retrofitted liquid nitrogen dewar inherited from the Keck Interferometer (Figure \ref{fig:zygo-periscope}), with the temperature of each optic stabilizing at SCALES' operating temperature of $<$100 K for at least an hour. This monitoring was possible only with flat optics, as the powered optics would have required an optical setup that the dewar did not have space for; however, powered optics were still thermally cycled inside the dewar, just without monitoring measurements being taken throughout. After the dewar was allowed to return to ambient temperature, the optic was then remeasured directly to detect any plastic deformation. Any optic that showed any change in surface form underwent at least one additional cryocycle. By combining before-and-after comparative measurements and monitoring throughout cryocycles, we are able to characterize both plastic and elastic deformation of each optic. The RMS wavefront error from surface figure for all optics and surface figure stability over repeat cryocycles is summarized in Table \ref{tab:summary}.

\begin{figure}[H]
\captionsetup[subfigure]{position=b}
\centering
    \subcaptionbox{Surface figure of OAP1.1 as received from son-x, before any cold testing.}{\includegraphics[width=0.42\linewidth]{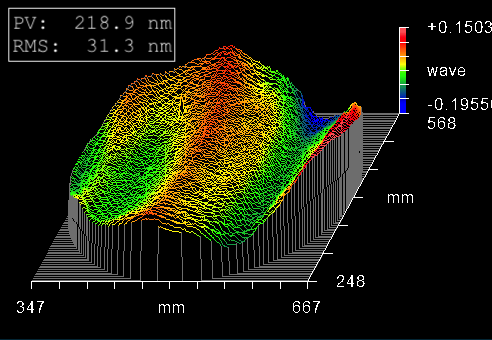}}
    \subcaptionbox{Surface figure of OAP1.1 after first complete cryocycle.}{\includegraphics[width=0.42\linewidth]{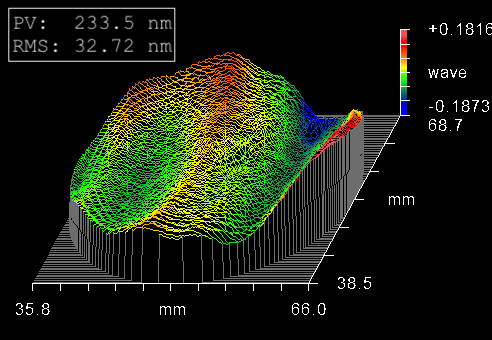}}
\smallskip
\caption{Surface form stability over one cryocycle of OAP1.1. Both measurements are taken at ambient temperature.}
\label{fig:cryocycle-post}
\end{figure}

\begin{figure}[H]
     \centering
     \begin{subfigure}[b]{0.49\textwidth}
         \centering
         \includegraphics[width=\textwidth]{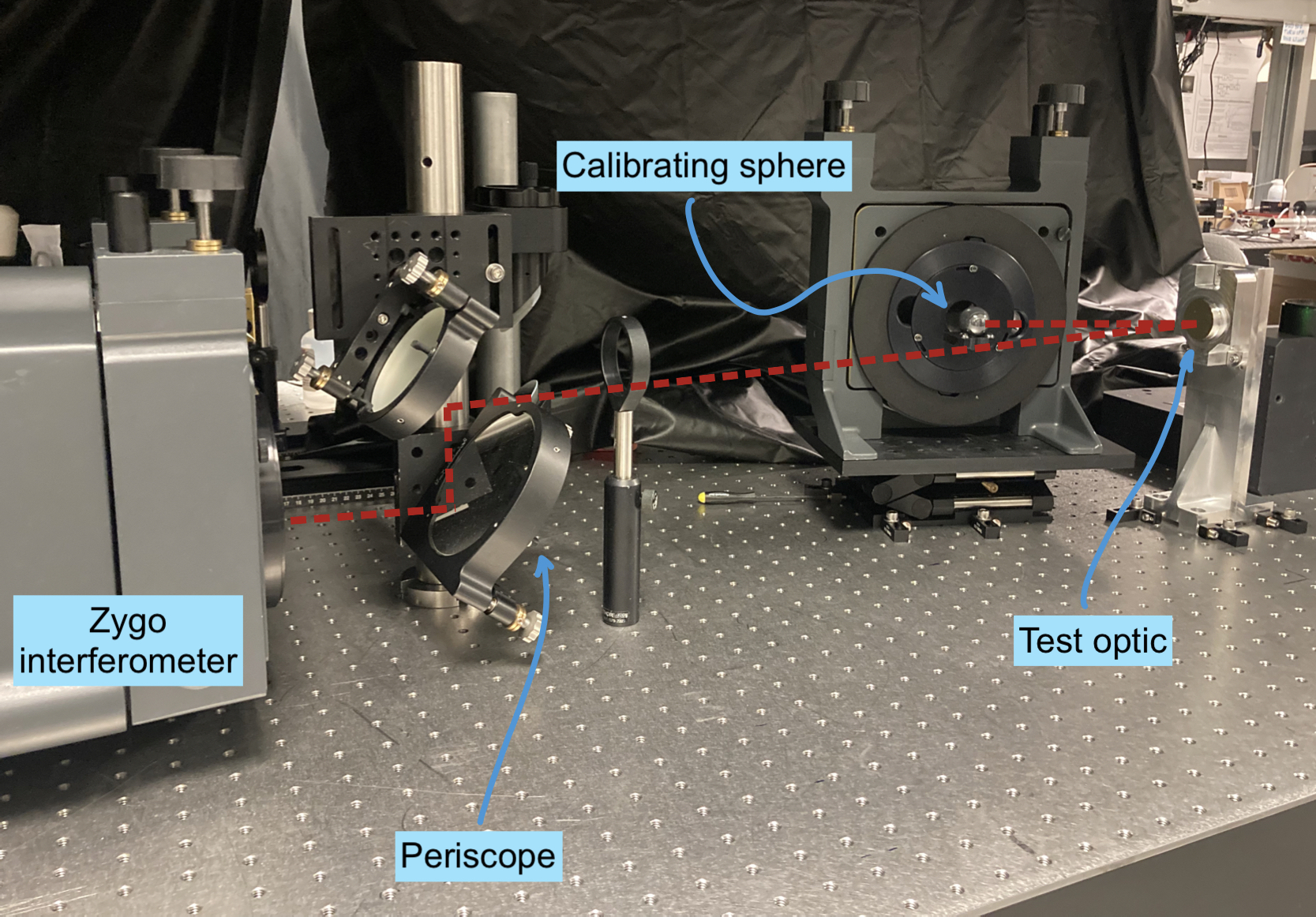}
         \caption{Ambient measurement setup}
         \label{fig:zygo-direct}
     \end{subfigure}
     \hfill
     \begin{subfigure}[b]{0.49\textwidth}
         \centering
         \includegraphics[width=\textwidth]{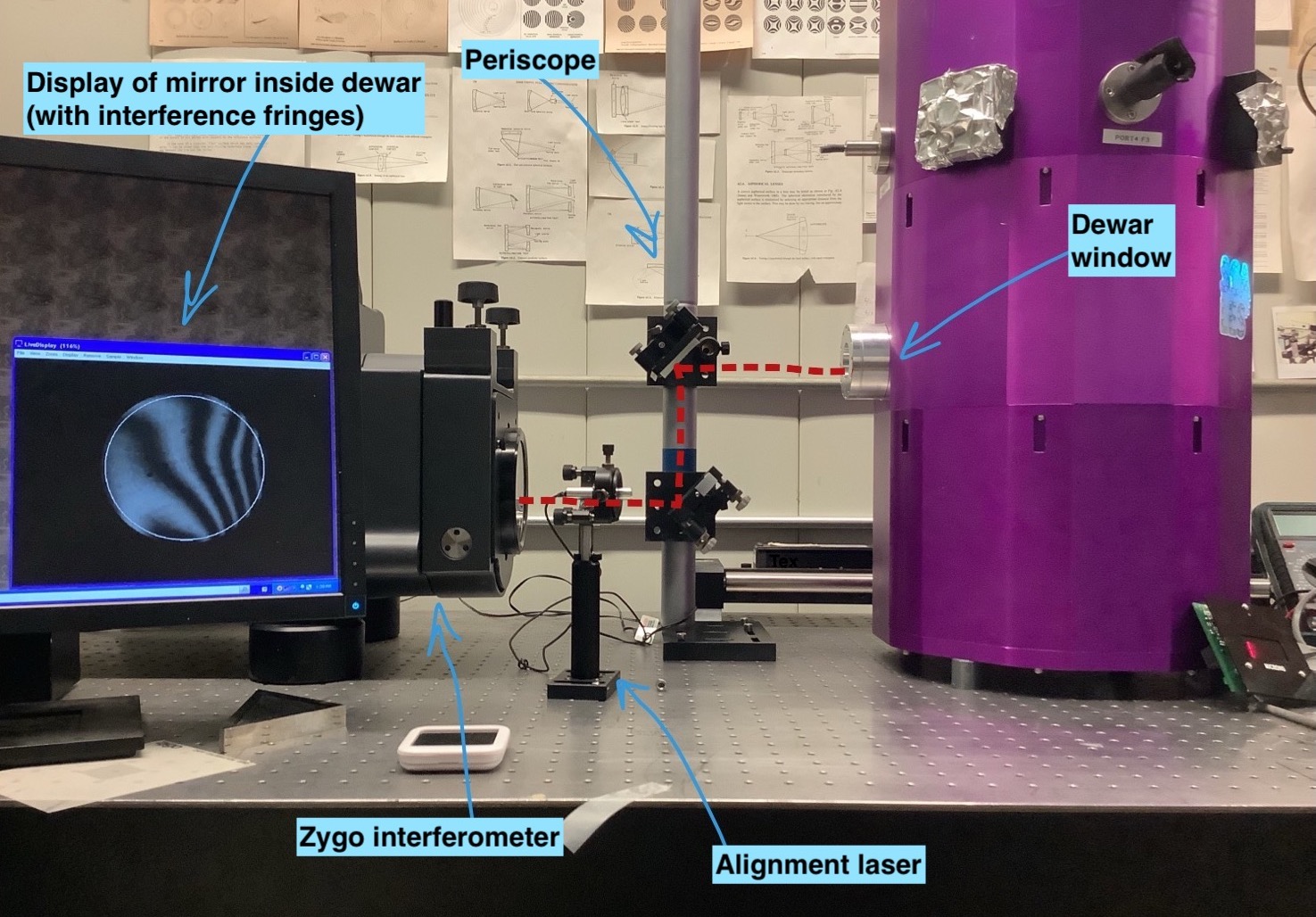}
         \caption{Cryogenic measurement setup}
         \label{fig:zygo-periscope}
     \end{subfigure}
        \caption{Test setups for measuring low-spatial frequency error with the Zygo interferometer. Left: setup used to characterize powered optics warm and outside vacuum (calibration sphere is omitted for flat optics). Right: setup used to monitor flat optics during cryocycling.}
        \label{fig:test-zygo}
\end{figure}

Figure \ref{fig:cryocycle-post} shows measurements of OAP1.1 before (left) and after (right) it underwent one thermal cycle inside the cryostat. While the clear aperture of the optic shows a change in surface form of $\leq$1.4 nm RMS, the surface form within the 15 mm beam footprint showed no change, and we decided to accept this optic without going through a second cryocycle. Additionally, this optic was very difficult to align, and the error bar on our alignment is approximately the same magnitude as the observed change.

\subsection{Surface roughness}
\label{subsect:roughness}

Surface roughness refers to the texture of the optical surface, measured as the RMS of the surface such that the distribution of surface irregularities is Gaussian about the mean. This high spatial frequency error scatters light at high angles out of the beam path, reducing the efficiency of the optical system and degrading overall performance. 

\begin{figure}[H]
\begin{minipage}[c]{0.49\linewidth}
\includegraphics[width=\linewidth]{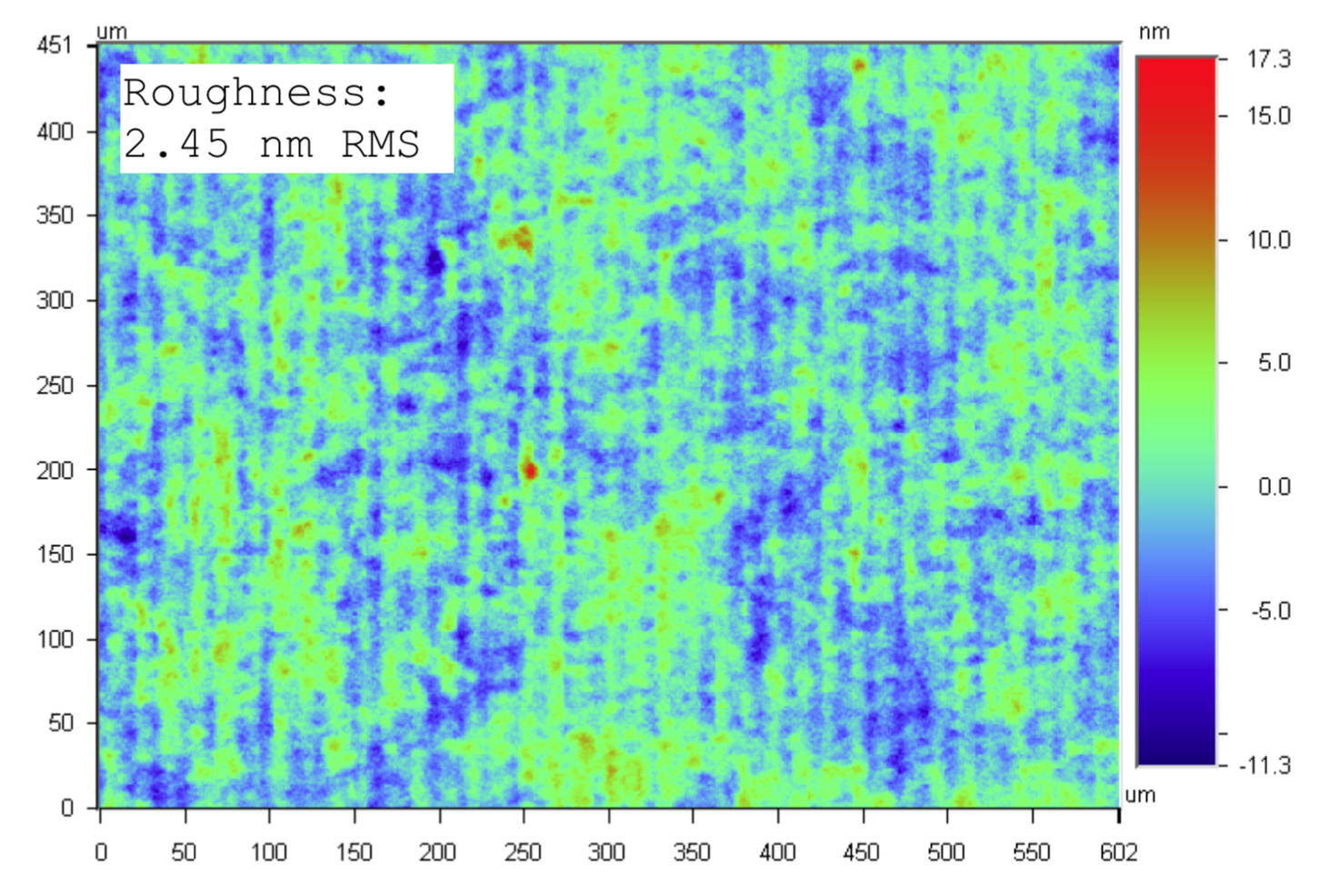}
\caption{Roughness measurement of OAP1.1. Tooling grooves from diamond turning are \\visible running vertically across the optic. \vspace{14pt}}
\label{fig:roughness}
\end{minipage}
\hfill
\begin{minipage}[c]{0.49\linewidth}
\includegraphics[width=\linewidth]{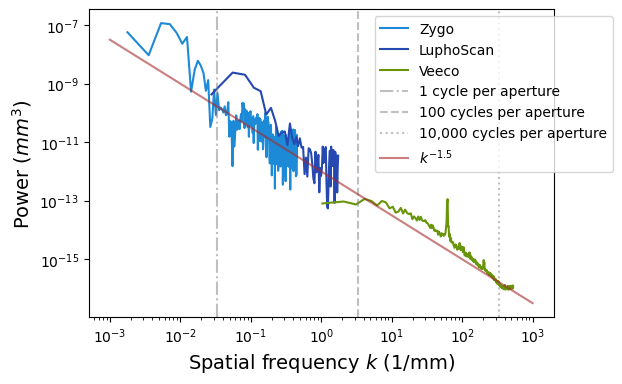}
\caption{PSD of OAP1.1. A power law with index -1.5 is fit by eye to the PSD. The tooling grooves cause a spike in the PSD at 60 mm$^{-1}$.}
\label{fig:psd}
\end{minipage}%
\end{figure}

The surface roughness of each optic is measured with a Veeco Wyko white light interferometric profiler, which is sensitive to spatial frequencies between 1-530 mm$^{-1}$. A 602 x 451 $\mu m$ measurement is taken of an unblemished spot within the clear aperture of each optic. Figure \ref{fig:roughness} shows an example measurement, and final values are reported in Table \ref{tab:summary}. Of the two optics measured so far, both meet the surface roughness specification of $<$3 nm RMS.

\subsection{Dimensional instability of RSA 443 under cryogenic conditions}
\label{subsect:failure}

The first optic received for testing was FM3 (see Figure \ref{fig:optics} for an optical diagram of SCALES), a flat mirror which was to be mounted to a cryogenic piezoelectric tip-tilt stage and used to steer between the low- and medium-resolution modes of the SCALES IFU. To avoid a CTE differential between the optic and the ceramic stage of the tip-tilt mechanism, the mirror substrate was initially chosen to be RSA 443 (where all other optics are RSA 6061), a polycrystalline matrix of 40\% silicon and 60\% aluminum. The substrate is coated with electroless nickel (nickel-phosphorous) before machining, since RSA 443 is too soft to yield excellent surface finish after diamond turning \cite{Mkoko2015AspectsAluminium}. The nickel coating and aluminum substrate have closely matching CTEs, and bimetallic bending was not expected.

We stepped through the measurement procedure for surface figure and surface roughness outlined in Subsections \ref{subsect:figure} and \ref{subsect:roughness}. Initial measurements of the mirror satisfied tolerances, with a measured surface roughness of 1.63 nm RMS and surface figure error of 23.3 nm RMS (Figure \ref{fig:roughness}). However, cryocycling the mirror inside the dewar caused plastic deformation of the mirror, and the surface figure worsened with each cryocycle. The surface figure creep over two full cryocycles is shown in Figure \ref{fig:creep}; the slight power and astigmatism (which are artifacts from manufacturing) visible in initial warm measurements intensify with each thermal cycle.

\begin{figure}[!ht]
\captionsetup[subfigure]{position=b}
\centering
    \subcaptionbox{Cryocycle 1
        }{\includegraphics[width=.40\linewidth]{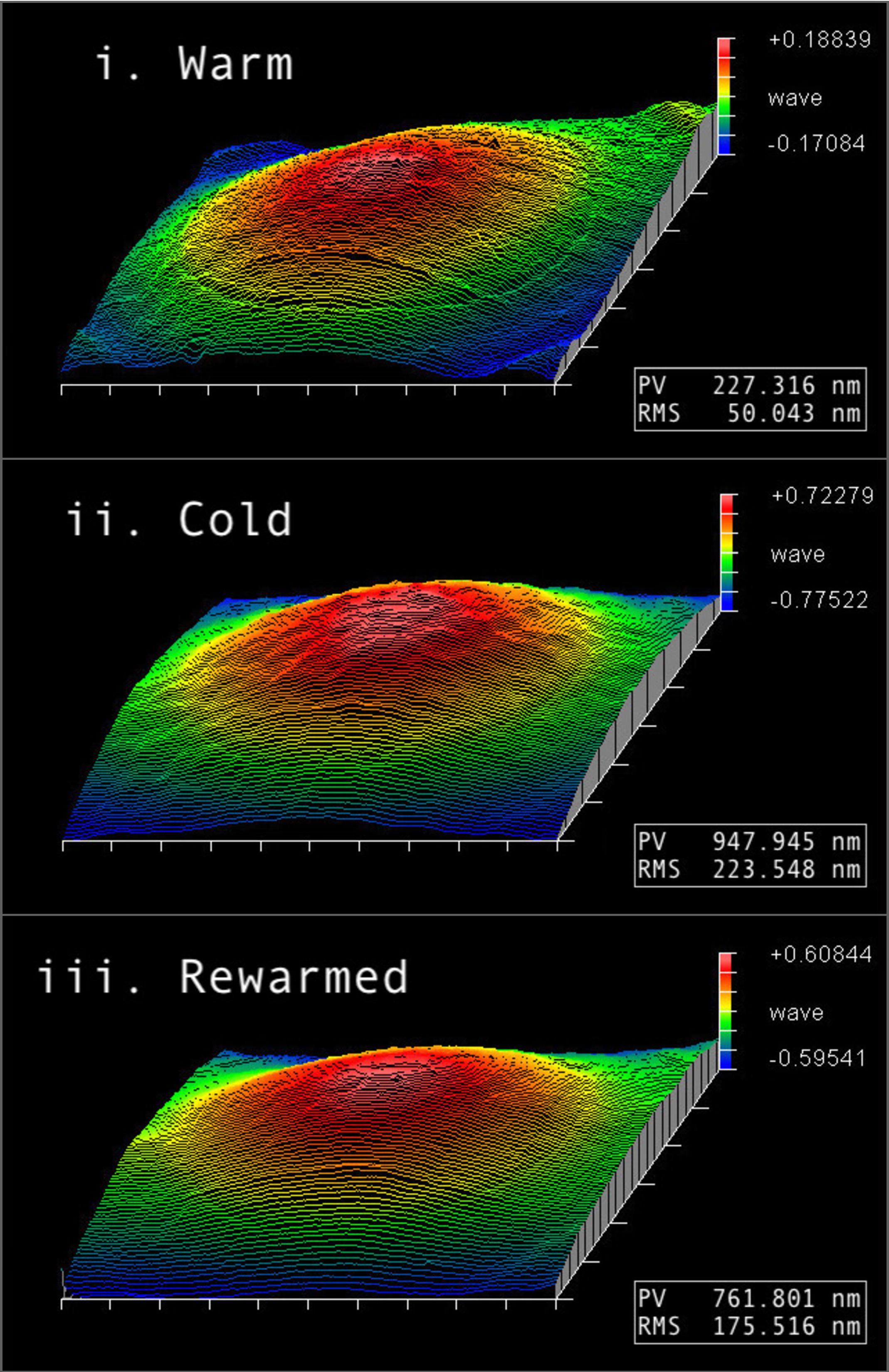}}
    \subcaptionbox{Cryocycle 2
        }{\includegraphics[width=.40\linewidth]{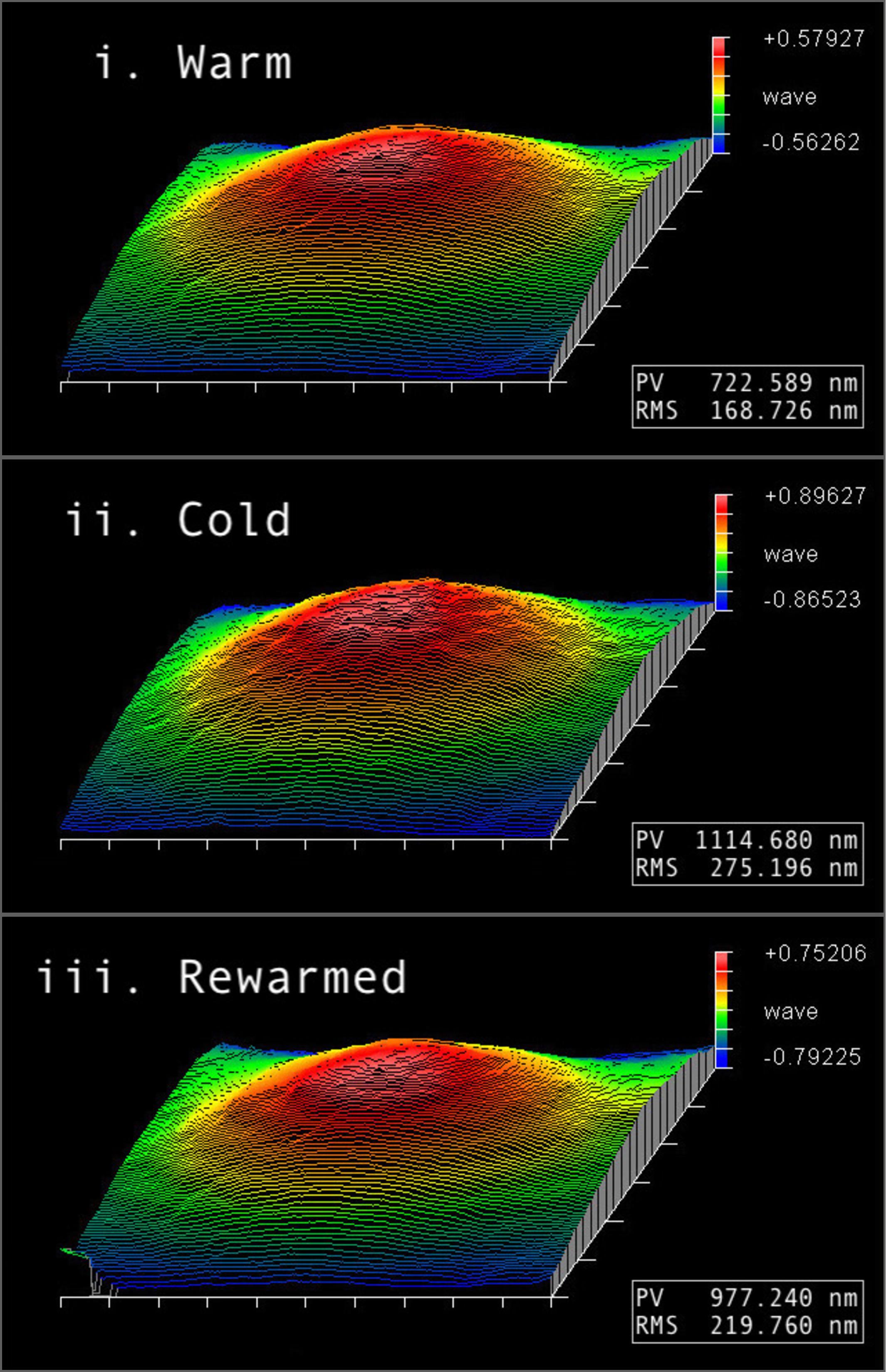}}
\caption{Measurements of the failed version of FM3 through two thermal cycles. All measurements are taken inside the cryostat at and vacuum pressure. Warm measurements are taken at ambient temperature, cold measurements are taken once the optic has settled near 77K for several hours, and rewarmed measurements are taken once the optic warms back up to ambient temperature. The optic has been removed from and replaced inside the dewar between the bottom left and top right panels, but has otherwise been subjected to no changes – the change in RMS can be attributed to differences in vacuum pressure affecting the curvature of the cryostat window.}
\label{fig:TTSM-cryocycle-during}
\end{figure} 

\begin{figure}[!ht]
\captionsetup[subfigure]{position=b}
\centering
    \subcaptionbox{Surface figure of optic as received from son-x.}{\includegraphics[width=.32\linewidth]{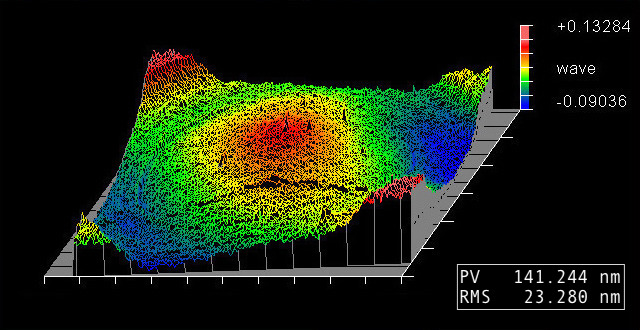}}
    \subcaptionbox{Surface figure of optic after first complete cryocycle.}{\includegraphics[width=.32\linewidth]{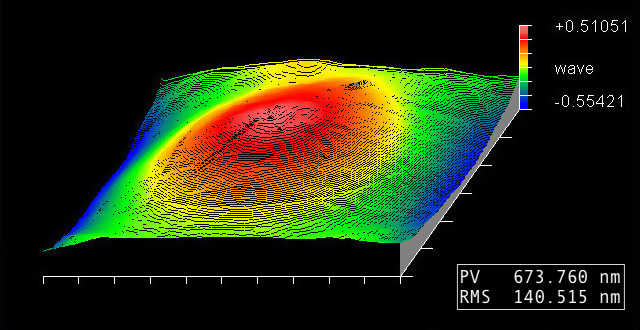}}
    \subcaptionbox{Surface figure of optic after second cryocycle.}{\includegraphics[width=.32\linewidth]{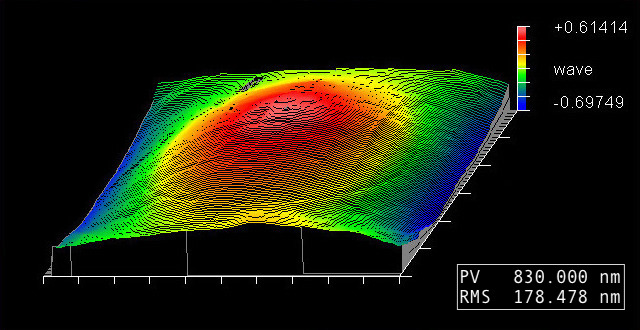}}
\caption{Surface form stability over multiple cryocycles of the failed RSA 443 optic (FM3). All measurements in this figure were taken warm and outside the dewar (Setup shown in Figure \ref{fig:zygo-direct}).}
\label{fig:TTSM-cryocycle-post}
\end{figure} 

The initial hope was that after a sufficiently high number of cryocycles, the material would stabilize and no longer experience plastic deformation, at which point the optic would be resurfaced and recoated to return its surface form to specifications. Since cryocycles in the liquid nitrogen dewar take about a week, and because plastic deformation post-cryocycle was our primary concern, we decided to rapidly cryocycle the mirror by suspending it over a bath of liquid nitrogen. Since the calculated radiative cooling timescale of the optic was on the order of weeks, we attached screws to the anchor points on the mirror and allowed these to dangle in the LN2 for thermal contact. To minimize condensation on the surface of the mirror, we flowed dry nitrogen gas across the face of the optic through the entire rapid cryocycle. Temperature sensors were anchored to various parts of the mirror to track cooling time and avoid dunking it completely. Surface form was remeasured after each rapid cryocycle (Figure \ref{fig:creep}). While plastic deformation asymptotically leveled off across 16 cryocycles, it did not fully stabilize, and any amount of creep was deemed unacceptable. With sufficient prior cryocycling, RSA 443 optics could be appropriate for certain cryogenic applications, though it is unknown whether complete dimensional stability is possible.

\begin{figure}[ht]
     \centering
     \begin{subfigure}[b]{0.49\textwidth}
         \centering
         \includegraphics[width=\textwidth]{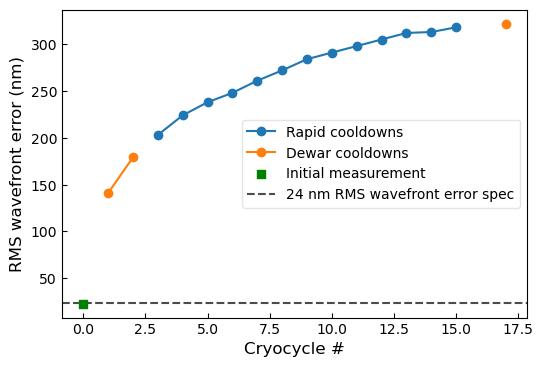}
         \caption{RMS wavefront error after each cooldown}
         \label{fig:creep-RMS}
     \end{subfigure}
     \hfill
     \begin{subfigure}[b]{0.49\textwidth}
         \centering
         \includegraphics[width=\textwidth]{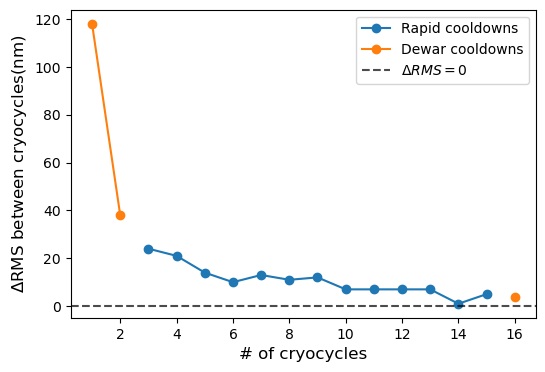}
         \caption{Change in RMS WFE between cooldowns}
         \label{fig:creep-deltaRMS}
     \end{subfigure}
        \caption{RMS figure error over 16 cryocycles, including both full multi-day cooldowns in a liquid nitrogen dewar (orange) and rapid cooldowns where the mirror was suspended over a bath of liquid nitrogen (blue). }
        \label{fig:creep}
\end{figure}

The culprit for this perpetual creep was eventually identified as the material properties of the mirror substrate, RSA 443 (though bimetallic bending from a CTE-mismatch between the optic substrate and coating is not ruled out). The lack of dimensional stability under thermal changes for a similar alloy has been reported in the literature by Caleta et al. 2017 \cite{Caleta2017DimensionalAlSi40} (Figure 5, Table 2). They found that samples of AlSi40, an alloy with similar composition to RSA 443, showed strong deformation during the first $\sim$6 thermal cycles from ambient to 77\degree K (-196\degree C). Further thermal cycling that reached minimum temperatures of 233\degree K (-40\degree C) showed dimensional creep that asymptotically leveled off (though never completely stabilized) after $\sim$20 cryocycles. Their findings included that AlSi40 alloys with finer particle sizes (like those yielded by the melt-spinning process involved in the manufacturing of RSA 443) tend to have poor dimensional stability, but that coarser-grained alloys show excellent stability even when subjected to cryogenic cycling. RSA 443 was also known to be dimensionally unstable \cite{RSPTechnology2022PrivateCommunication}, but this was not publicly well-documented. While the cause of plastic deformation of RSA 443 is not fully characterized, the elastic deformation is thought to be caused by a CTE mismatch between the aluminum matrix and silicon particles \cite{RSPTechnology2022PrivateCommunication}. 

Ultimately, we find that RSA 443 is an unacceptable substrate for this instrument. son-x remade the part using an RSA 6061 substrate, and the tip-tilt mechanism was partially re-manufactured by Physik Instruments to accommodate the change in material from RSA 443 to RSA 6061.


\section{Simulating the SCALES PSF and limitations on contrast performance}
\label{sect:contrast}


This work characterizing SCALES' WFE is motivated by impact on contrast performance. Using \texttt{poppy} (Physical Optics Propagation in PYthon) \cite{Perrin2016POPPY:PYthon}, we constructed a high-fidelity Fresnel physical optics propagation model that begins with the Keck primary and terminates at the focus placed on the lenslet array by the foreoptics. This assembly includes a representation of each existing optic, with the foreoptics represented by real measurements of WFE across spatial scales, and all other optics (Keck telescope and AO optics) are represented by unaberrated surfaces of specified size and optical power.

Representations of WFE from the SCALES foreoptics between 0.0018-70 mm$^{-1}$ are created from lab measurements and inserted into \texttt{poppy}. A low spatial frequency WFE component (0.0018-0.45 mm$^{-1}$) for each optic is built from Zygo measurements (Section \ref{subsect:figure}). A mid-high spatial frequency component is constructed using a power law model fit to the WFE power spectral density (PSD) constructed from all measurements; for OAP1.1 (the one optic where all measurements have been completed), we found a best fit power law index of $\alpha$ = -1.5 (see Figure \ref{fig:psd}). The upper limit of spatial frequency included in this model is bounded by the sampling of the \texttt{poppy} model, not by the sensitivity of our lab measurements, since the optical profiler used to take surface roughness measurements is sensitive to spatial frequencies from 1-530 mm$^{-1}$. However, neglecting high spatial frequency error in our model does not significantly change the resulting simulated PSF -- high spatial frequency error effectively lowers the throughput of an optical system, scattering light at wide angles out of the PSF. While this slightly degrades contrast performance, it does not shape the PSF. 

This simulation also considers the residual uncorrected wavefront error from the Keck AO system, which is quantified in the Keck AO Error Budget \cite{KAON1303}. We ignore error sources that vary quickly in time and thus average out over realistic exposure times (we specifically neglect atmospheric fitting error, bandwidth error, high-order measurement error, and high-order aliasing error). Power-law representations with a negative index of 3 (by convention) of the residual wavefront error from the current Keck AO system (189 nm RMS) and a future Keck AO upgrade, HAKA (High order Advanced Keck Adaptive optics, 162 nm RMS) are incorporated into the simulation. 

HAKA introduces a 60x60 actuator high-order deformable mirror into the Keck AO system (compared to the current 20x20 actuator DM), and involves upgrading the real time controller (RTC) and the Shack-Hartmann wavefront sensor. HAKA is expected to be commissioned at around the same time as SCALES, which is slated for delivery in summer of 2025. Improved AO performance with HAKA will primarily drive down error occurring on short timescales (with greatest improvements in atmospheric error and bandwidth error) and at close separations, which this simulation is agnostic to. Since this work neglects the improvements in short-timescale error terms, the primary difference between this representation of current and future Keck AO WFE lies in uncorrectable telescope and AO system aberrations, which will be additionally driven down by improved phasing of the Keck primary mirror segments \cite{Ragland2022ResidualTelescopes, vanKooten2022On-skySensor} and more advanced controls that improve the nulling of aberrations that originate within the AO system. \cite{Bos2021FastObservatory} Thus the true difference in science performance between current and future Keck AO will be more pronounced than what is shown here. 

\begin{figure}[bp!]
    \centering
    \includegraphics[width=0.87\textwidth]{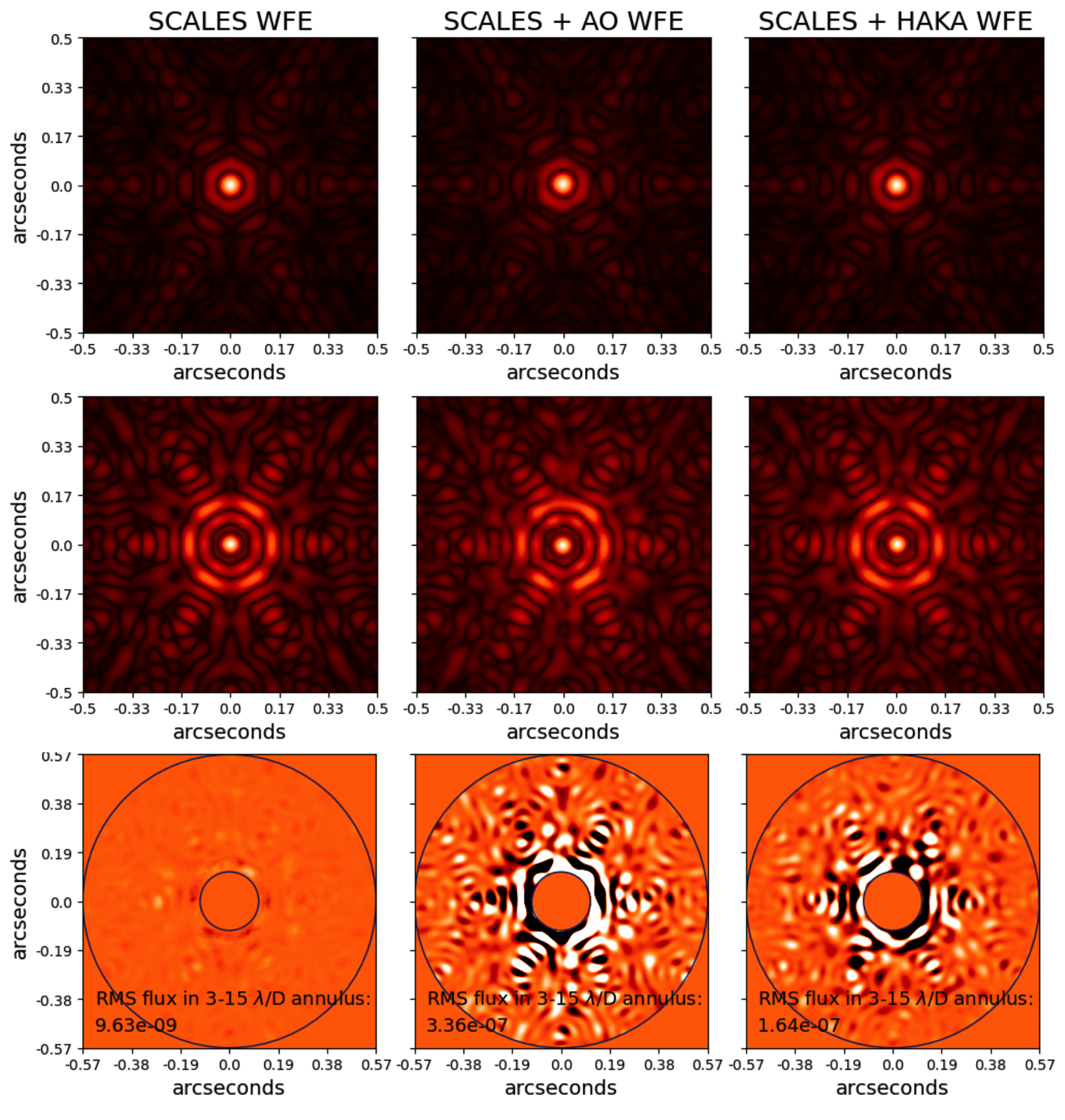}
    \caption{SCALES PSF simulated in \texttt{poppy} at $\lambda = 2\mu m$, showing three wavefront error cases: only considering WFE from the SCALES foreoptics (left column); considering WFE from SCALES and 189 nm RMS of uncorrected WFE from the current Keck AO system (center column); and considering WFE from SCALES and 162 nm RMS WFE from HAKA, an upcoming upgrade to the Keck AO system. The top panel shows the PSF placed by the foreoptics on the lenslet array (see Figure \ref{fig:optics}) with no coronagraph. The middle panel shows the PSF with a Lyot-style coronagraph with a 6$\lambda$/D circular occulter and a Lyot stop designed to match the shape of the Keck pupil \cite{Li2021Cold-StopObservatory}. The bottom panel shows a rough approximation of reference differential imaging (RDI), where five realizations of the simulated PSF are averaged together and subtracted from one simulated PSF to reveal a starlight-subtracted PSF. A 3-15 $\lambda$/D annulus is overlaid onto the subtracted PSFs, and the RMS fractional flux contained in that annulus is annotated. Each horizontal panel shares the same color scale.}
    \label{fig:psf}
\end{figure}

The modeled SCALES PSF is shown in Figure \ref{fig:psf}. To compare the impact of different sources of WFE on contrast, three wavefront error sources are considered:
\begin{itemize}
    \item \textbf{SCALES}: The Keck telescope and AO system are assumed to be perfect, and only measured WFE from the SCALES foreoptics is included.
    \item \textbf{SCALES + AO}: WFE from SCALES foreoptics and the current Keck AO system are both considered. The WFE contribution from the current AO system for a 5th magnitude natural guide star, neglecting quickly varying error terms, is 189 nm RMS. 
    \item \textbf{SCALES + HAKA}: WFE from SCALES foreoptics and the proposed future Keck AO upgrade, HAKA (High order All-sky Keck Adaptive optics), are both considered. HAKA integrates an additional high-order deformable mirror into the AO bench, and the projected residual WFE for a 5th magnitude natural guide star (neglecting quickly varying error terms) is decreased to 162 nm RMS. 
\end{itemize}



While \texttt{poppy} does not have the option to directly model the vector vortex coronagraph that will be integrated into SCALES, we insert a a Lyot-style coronagraph with a 6$\lambda$/D diameter circular top-hat occulter (3$\lambda$/D inner working angle) and a Lyot stop designed to match the shape of the Keck pupil \cite{Li2021Cold-StopObservatory} to approximate the starlight suppression of a more sophisticated coronagraph. The top panel of Figure \ref{fig:psf} shows the unsuppressed PSF, and the middle panel shows the PSF with the majority of on-axis light blocked by the coronagraph.

Directly imaged exoplanets are often too faint to be visible even with starlight suppression from a coronagraph, and are only revealed after post-processing. To create a rough approximation of reference differential imaging (RDI), a template PSF is created by averaging together five simulated PSFs (which show stochastic differences due to the way WFE is represented in the model), and this template is then subtracted from a single realization of the simulation to reveal a starlight-subtracted PSF, shown in the bottom panel.

The starlight-subtracted images outside of a 3-15 $\lambda$/D annulus are masked. These subtracted PSFs are annotated with the RMS flux contained within that annulus, which is where aberrations from mid-spatial resolution WFE become apparent, and where SCALES will look for exoplanets. The flux values in the final PSF are not in physical units, but represent fractional brightness compared to the entrance pupil, which has an integrated flux of 1. The subtracted SCALES WFE PSF has an RMS flux of 9.63e-9; the SCALES + Keck AO WFE PSF has an RMS flux of 3.36e-7; and the SCALES + HAKA WFE PSF has an RMS flux of 1.64e-7. These subtracted PSFs demonstrate that the speckle noise floor is dominated by uncorrected wavefront error from the Keck AO system, not WFE contributions from SCALES foreoptics, even when the near-future HAKA upgrade is considered.

\begin{figure}[bp!]
    \centering
    \includegraphics[width=0.9\textwidth]{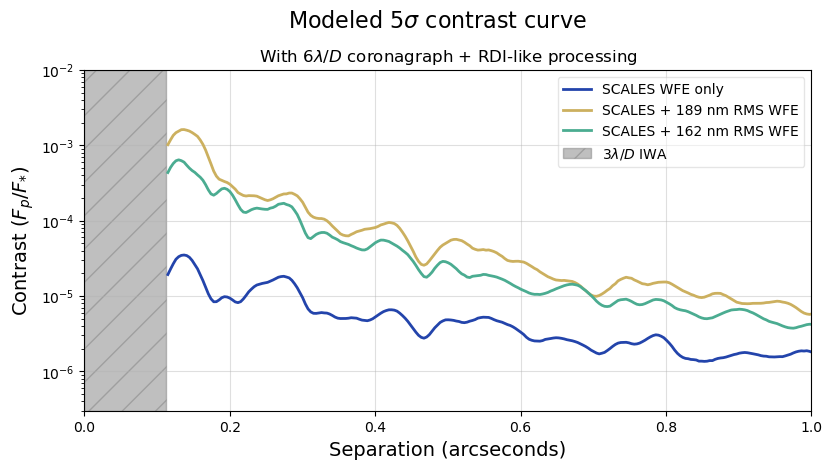}
    \caption{Simulated 5$\sigma$ contrast for three wavefront error cases: SCALES foreoptics only; SCALES foreoptics plus 189 nm RMS of uncorrected WFE representing the summed WFE of the current Keck AO system; and SCALES foreoptics plus 162 nm RMS of uncorrected WFE representing an upcoming upgrade to the Keck AO system, HAKA. All AO WFE estimates are taken from KAON 1303 \cite{KAON1303}, and assume a 5th magnitude natural guide star. This simulation neglects error terms that vary on short timescales and thus average down during realistic exposure times, and thus is dominated by uncorrectable static and dynamic telescope aberrations. These contrast curves are not intended to be contrast predictions, but instead demonstrate that SCALES' contrast is not limited by internal optics, but by performance of the AO system. For a dedicated simulation of SCALES + HAKA contrast performance, see Sallum et al. 2023 \cite{Sallum+2023}.}
    \label{fig:contrast}
\end{figure}

The contrast curves in Figure \ref{fig:contrast} are derived from the suppressed PSFs in the bottom panel of Figure \ref{fig:psf}. The suppressed PSF is convolved with a Gaussian with a FWHM of 1.22$\lambda$/D. 1$\sigma$ contrast is calculated by taking the azimuthal standard deviation at a range of separations from the center of the Gaussian-smoothed PSF and normalizing by the peak brightness of an unsuppressed simulated PSF. The 3$\lambda/D$ inner working angle of the coronagraph included in the \texttt{poppy} model is masked.

Predicting instrument contrast is notoriously tricky, and developing an accurate contrast prediction is not the purpose of this work. It is also difficult to quantify whether this contrast curve is optimistic or pessimistic, though we note a number of generalizations we include in our modeling. The coronagraph included in the \texttt{poppy} model is a hard-edged circular occulter, which successfully suppresses on-axis light but introduces diffraction effects around the hard edges of the mask. Other more sophisticated coronagraphs avoid this issue and show better contrast performance, including the vector vortex coronagraph that will be installed in SCALES. Additionally, our modeling takes into account telescope and instrument optics, but does not consider any realistic observing scenario. Normally, contrast can be deepened by extending integration time to collect more signal, and by doing some combination of spectral differential imaging (SDI), angular differential imaging (ADI), and reference differential imaging (RDI). We do a quick approximation of RDI by subtracting a template PSF (built from five realizations of the PSF simulation averaged together) from a single simulated PSF, but a real observation would combine several post-processing techniques. For a SCALES contrast curve that approximates a real observation, see Sallum et al 2023. \cite{Sallum+2023}  


The measurements and simulations presented in this work are not tailored specifically for accurate contrast prediction, which requires much more intensive modeling. Instead, this is meant to comparatively demonstrate the limiting wavefront error sources contributing to contrast performance. Figure \ref{fig:psf} and \ref{fig:contrast} demonstrate that SCALES' contrast performance is \ dominated by wavefront error from the Keck AO system, and that the wavefront error imparted by SCALES foreoptics is not the limiting factor in determining contrast performance. 



\section{Summary of results}
\label{sect:summary}

\begin{itemize}
    \item Diamond-turned aluminum-substrate optics demonstrate excellent WFE (a range of 1.5-3.0 nm RMS surface roughness and 16-29 nm RMS reflected WFE) appropriate for use in astronomical instrumentation.
    \item The only RSA 6061 optic tested cryogenically thus far, OAP1.1, shows an upper bound of $\leq$1.4 nm RMS of plastic deformation. We were unable to monitor elastic deformation during cryocycling due to the size constraints of our cryostat. Future updates will present typical values for elastic and plastic deformation during cryocycling for the full suite of aluminum foreoptics. 
    \item RSA 443 is inappropriate for use in the SCALES instrument. The first cryocycle of an RSA 443 optic showed plastic deformation of 117 nm RMS and elastic deformation of 173 nm RMS; after 16 thermal cycles, plastic deformation asymptotically decreased but never fell below 1 nm RMS of creep per thermal cycle. 
    \item Contrast performance of the SCALES instrument will be limited by the performance of the Keck AO system, not by WFE introduced by internal instrument optics
    \item All measurements are presented in Table \ref{tab:summary}.
\end{itemize}

\begin{table}[H]
\centering
\renewcommand{\arraystretch}{1.15}
\caption {\label{tab:summary}Summary of measurements. Missing measurements will be filled in as finished optics are shipped to UCSC.}
\begin{tabularx}{\textwidth}{ Y | Y | c c c }
    \hline\hline
     Optic &  Surface roughness WFE (nm RMS) & \multicolumn{3}{Y}{Surface figure WFE (nm RMS)}  \\
     \hline
     && 300K & 77K & $\Delta$RMS$^a$ \\
     \hline\hline
        OAP1.1 & 2.5 & 27.6 & -- & 1.4 \\
        OAP1.2 & -- & -- & -- & -- \\
        OAE & -- & 16$^b$ & -- & --\\
        FM1 & 3.12$^b$ & 20$^b$ & -- & -- \\
        FM2 & -- & -- & -- & -- \\
        FM3 (RSA 443) & 1.63 & 23.3 & 140.5 & 117.2 \\
        FM3 (RSA 6061) & -- & 18$^b$ & -- & -- \\
        FM4 & -- & -- & -- & -- \\
        FM5 & -- & -- & -- & -- \\
    \hline\hline
\end{tabularx}

\raggedright
\smallskip

$^{\mathrm{a}}\Delta$RMS refers to the difference in RMS surface figure between the first and second cryocycles, i.e. plastic deformation, with both measurements being taken inside the cryostat at 77K. \\
$^{\mathrm{b}}$ These measurements were taken by son-x before the optics were coated, and will be remeasured at UCSC once they are coated and shipped.\\
\end{table}

\section{Future work}
\label{sect:future}

At the time of publication, we have received and tested OAP1.1 and an initial version of FM3 that failed during testing. FM3 was remade with a different substrate, but for the sake of scheduling was integrated directly into the tip-tilt mechanism before running the tests described in this work. We will repeat the outlined tests on these six remaining optics (FM1, FM2, FM4, FM5, OAP1.2, and OAE) once they're received, and we will complete testing of FM3 alongside testing of the tip-tilt stage.

Tooling marks from single-point diamond turning have been known to cause diffraction effects\cite{Wu2021DiffractiveTurning}. We intend to measure IR diffraction behavior of all optics once received using a ThorLabs SLS202L 0.45 - 5.5 $\mu m$ broadband source and a ThorLabs 180C 2.9 - 5.5 $\mu m$ power sensor. This measurement was carried out on the RSA 443 version of FM3 that later failed in testing, but subsequent optics were manufactured with a different groove spacing by request of the SCALES team, and thus the diffraction behavior of other optics is expected to be different.


\section{Acknowledgements}

We are grateful to the Heising-Simons Foundation, the Alfred P. Sloan Foundation, and the Mt. Cuba Astronomical Foundation for their generous support of our efforts. This project also benefited from work conducted under NSF Grant 2216481 and the NSF Graduate Research Fellowship Program. We thank the generous individual donors who have supported the SCALES project and our student researchers. The specific work conducted in this paper was made possible by Barton Robinson, Philip Rice and the Webster Fellowship program. We thank Rebecca Jensen-Clem and Peter Wizinowich for their technical correspondence.

\appendix

\bibliography{references}   
\bibliographystyle{spiejour}   

\listoffigures
\listoftables

\end{spacing}
\end{document}